\documentclass{PoS}
\usepackage{amssymb}

\newcommand{\simprop}{\mathrel{\raise.25ex\hbox{$\propto$}\kern-0.78em\lower.9ex\hbox{$\sim$}}}
\newcommand{\approxprop}{\mathrel{\raise.25ex\hbox{$\propto$}\kern-0.78em\lower.9ex\hbox{$\approx$}}}
\newcommand{\criteq}{\mathrel{\kern0.5em\raise1.41ex\hbox{$!$}\kern-0.5em\hbox{$=$}}}
\newcommand{\critsim}{\mathrel{\kern0.5em\raise1.41ex\hbox{$!$}\kern-0.5em\hbox{$\sim$}}}
\newcommand{\critlsim}{\mathrel{\kern0.5em\raise1.72ex\hbox{$!$}\kern-0.55em\hbox{$\lesssim$}}}

\title{Origin of ultra-high energy cosmic rays:\\ 
The cosmological manifesto}

\ShortTitle{Origin of UHECR: Cosmological Manifesto}

\author{\speaker{J\"org P. Rachen}$^{\,1,2}$ for the \textit{CCCP}\\
        $^1$ Astrophysical Institute, Vrije Universiteit Brussel, Belgium\\
        $^2$ Department of Astrophysics/IMAPP, Radboud University, Nijmegen, the Netherlands\\
        E-mail: \email{jorg.paul.rachen@vub.be}}

\abstract{A spectre is haunting physics---the spectre of non-thermal processes. Manifesting itself in astroparticle physics, all classical disciplines of physics and astronomy have entered into a holy alliance to exorcise this spectre: While astronomers try to destroy it with ignorance, convinced that only thermal processes are of importance to the dynamics of the Universe, particle physicists obscure it in the clouds of mystified admiration, simply because the Universe works on energy scales which are far above of their own technical and economical capabilities. It is high time to understand the energy scales of astroparticle physics in those of the Universe itself, by linking the existence of their basic essence, cosmic rays, via a universal mechanism to the ground reservoir of all non-thermal energy, gravitation. By this we meet the nursery tale of the enigmatic, mysterious sources of cosmic rays with a set of numbers naturally derived from cosmological parameters, which manifest themselves in a large range of astrophysical objects that stand in rank and file along a diagonal in the famous Hillas-plot, and in fact only have in common that they efficiently convert gravitational energy. While this manifesto, promoting the cosmological revolution in astroparticle physics, does not release us from detailed modelling of cosmic ray and multimessenger sources in order to understand all the details delivered by current experiments, it does help to clear up our view on the problem and to avoid some common fallacies and misconceptions.}

\FullConference{36th International Cosmic Ray Conference -ICRC2019-\\
		July 24th - August 1st, 2019\\
		Madison, WI, U.S.A.}

\begin{document}

\begin{footnotesize}

\section*{Preface to the ICRC2019 edition}

\noindent This manifesto differs significantly from usual, bourgeois science papers in that it presents a different way of thinking about astroparticle physics rather than presenting detailed results within the bourgeois paradigm. The Central Committee of the Cosmological Party (\textit{CCCP}\/) has therefore decided to use common mathematical symbols in a slightly different, more revolutionary meaning as follows. 

Let $A$ and $B$ be some comparable physical quantities. We will then write $A\simeq B$ to express that they approximately agree within the precision given. Otherwise we define
\[
    \begin{array}{lcr@{\qquad\Leftrightarrow\qquad}lcl@{\qquad}l}
        A &\approx& B &  \lg(A) &\in& [\,\lg(B) \pm \frac12\,] & \rm (speak: within\ order\ of\ magnitude)\\
        A &\sim& B &  \lg(A) &\in& [\,\lg(B) \pm 1\,] & \rm (speak: around\ order\ of\ magnitude)  
    \end{array}\nonumber
\]
Whenever these symbols are combined with symbols of inequality, they are combined in an ``and-sense''; so, e.g., $A\lesssim B$ can be read as ``smaller, but still around order of magnitude''. We also define a symbol $A \simprop (\,\approxprop\,)\,x$ to express that $A$ is approximately proportional to $x$, i.e., there is a constant $B$ so that $A\sim (\,\approx\,)\,B x$ for all $x$ in a given range. We furthermore use the notation $\tilde X$ for a quantity $X$ given in natural (Planck) units ($c=\hbar=G=1$). 

Apart from this, excessive referencing is avoided by decree of the \textit{CCCP}, as most of what is presented here is based on textbook knowledge. Instead, we generally refer to the textbooks of Peebles \cite{Peebles} and Dodelson \cite{Dodelson} for the subject of general cosmology and structure formation, where cosmological parameters are generally taken from the Planck legacy results \cite{PlanckLegacy}; to the fundamental work of Brandenburg and Subramanian on turbulence and magnetic field generation \cite{BS05}; to the review of Drury \cite{Drury} for Fermi acceleration, and to the textbooks of Berezinskii et al.~\cite{Berezinsky} and Stanev \cite{Stanev} for the physics of cosmic rays. Further references are given only where additional and relevant fundamental concepts or significant results are mentioned.

\end{footnotesize}

\section{Non-thermal processes, self-organized criticality and 1/\textit{f} laws}

\noindent Cosmic rays are a manifestation of a class of phenomena in nature called non-thermal processes. Unlike thermal processes, in which a system containing a large number of constituents distributes its total energy among them in individual interactions such that entropy is maximized (leading to the well known thermal distributions of particles and radiation), non-thermal processes are typically the result of \textit{collective processes}, in which subsystems of particles interact with each other. This drags a rat tail of other concepts behind it, like self-similarity and self-organization, which became increasingly popular in the physical and mathematical description of biological and social systems. An often observed feature in such systems are the so called $1/f$ spectra, commonly defined as 
\begin{equation}
\label{eq:1_over_f}
S(E) \equiv E\;\frac{dN}{dE} \propto E^{-(1 + \delta s)}  \qquad {\rm with} \quad | \delta s | < 0.5\;,
\end{equation}
where $S(E)$ is the energy spectrum (total energy per unit particle energy $E$), and $dN/dE$ the number spectrum (this is why the $1/f$ spectrum is often referred to as an ``$E^{-2}$ spectrum'' in cosmic ray physics). The most quoted attempt for an explanation of this phenomenon is the concept of self-organized criticality, which attributes it to a series of self-similar critical points, so that subsystems reaching this critical point are tend to ``slide'' through phase space until it is evenly occupied \cite{BTW88}.   

The two main essences realizing non-thermal processes in the universe are turbulence and magnetic fields.  Turbulence describes randomization of a large scale kinetic flow through a cascade of eddies, starting at the outer system scale $R$, and ending at the scale of the Debye length of a plasma below which the system thermalizes.  It is ignited wherever bulk kinetic energy in a tenuous plasma is converted into internal modes, for example at shocks, which randomize a fraction $\xi_{\rm turb} \simeq 1-1/r^2 \lessapprox 1$ of the bulk energy for a usual compression ratio $r\gtrapprox 3$, and follows an energy spectrum $E(k) \propto k^{-n_k}$ over wave-number $k$, with $n_k=\frac53$ corresponding to the most common case of Kolmogorov turbulence. If the plasma is magnetized and its Reynolds number is larger than a critical value ${\cal R} \approx 30$, turbulence can ignite dynamo action leading to an exponential growth of a random magnetic field that can be described by a spectrum of plasma waves with energy $u_B = [\delta B(k)]^2/8\pi \propto k^{-n_k}$ for large $k$. The timescale for establishing these turbulence spectra is of the order of the turnover time of the largest eddies, and the total magnetic energy density saturates at $u_B = \xi_{\rm dyn} u_{\rm turb}$ with $\xi_{\rm dyn} \approx 0.1$, so that we can define an efficiency for magnetic field generation as $\xi_B \equiv \xi_{\rm turb}\xi_{\rm dyn} \sim 0.1$.   

Having turbulence and magnetic fields established, the generation of cosmic rays is an inevitable consequence. Plasma waves in a turbulent magnetic field move along the field lines with characteristic velocities $v_{\rm A}$, called the Alfv\'en velocity, which is related to the characteristic kinetic velocity $v$ in the plasma as $v_{\rm A} = \sqrt{\xi_B}\,v$. These plasma waves can efficiently scatter particles gyrating in the magnetic field on radii comparable to the wavelength, igniting a stochastic gain of energy of these particle, as first realized by Comrade Fermi in 1949. It was later realized that in the vicinity of shock waves this \textit{Fermi-acceleration} becomes a scale-free, self-similar process, producing $1/f$ spectra of energetic particles with
\begin{equation}
\label{eq:delta_s_Fermi}
\delta s = \frac{3 \delta\gamma + 8 {\cal M}^{-2}}{2 - 2 {\cal M}^{-2}}
\end{equation}
where $\delta\gamma = \gamma - \frac53$ is the deviation of the adiabatic index from the ideal non-relativistic case, and $\cal M$ is the Mach number of the shock. As $-\frac13 < \delta\gamma \le 0$ we get $|\delta s| < 0.5$ for ${\cal M} > 3$. 

The most fundamental $1/f$ laws however, are found in the most fundamental scale-free force in nature: gravitation. So is the (truncated) Press-Schechter function, i.e., distribution of dark matter halo masses $M_{\rm h}$ formed in cosmological structure formation  
\begin{equation}
\label{eq:PressSchechter}
\frac{dN_{\rm h}}{d\ln M_{\rm h}} \;\;=\;\; M_{\rm h}\,\frac{dN_{\rm h}}{dM_{\rm h}} \;\;\approx\;\; A(n)\;\frac{\rho_c \Omega_m}{M} \left(\frac{M}{M_*}\right)^{\!\!\!\frac{n+3}{6}} \qquad{\rm for}\quad M<M_*\;,
\end{equation}
where $\rho_c = 3 H_0^2 / 8 \pi G$ is the critical density of the Universe evaluated for the current value of the Hubble parameter, $H_0\simeq 68\,\rm km\,s^{-1}\,Mpc^{-1}$, $\Omega_m \simeq 0.3$ the matter density parameter, and the normalization factor $A(n) = (n+3)/6$ for $n\neq -3$  and $A(-3) \simeq 1/\ln \widehat M_*$ with $\widehat M_*$ being the largest collapsing mass $M_*$ in units of the smallest ones. This is a $1/f$ law with $\delta s = -(n+3)/6$, originally derived by the concept of self-similar clustering with $n \simeq 0$ taking the value of the spectral index of primordial fluctuations. Detailed simulations have confirmed the self-similar, scale-free behavior of structure formation, but with effective values of $-3 \le n \le -1$ dependent on cosmological epoch, leading to $-\frac13 \le \delta s \le 0$. 

We now get to the crux of the revolution: All non-thermal processes can be understood on the base of the concept on self-organized criticality, meaning that the system is not governed by a a particular scale, but by a constant relation of scales, usually expressed in a dimensionless number. For structure formation, we can define the parameter
\begin{equation}
\label{eq:varsigma}
    \varsigma \;\;\equiv\;\; \frac{R c^2}{G M} \equiv \frac{c^2}{\phi_G} \;\;\criteq\;\; \frac{k T}{m c^2} \equiv \frac{1}{c^2}\left[\frac{d\sigma^2_v}{d\log R}\right]^{-1} 
    \;\critlsim\;\; \frac{2\Omega_m^{1.2}\,H_0^2\,R_p^2\sigma_p^2}{3c^2} \;\;\approx\;\;  10^7\; 
\end{equation}
which takes a similar role in structure formation as the (magnetic) Reynolds number does in (magnetic) turbulence, or the Mach number and adiabatic index do in first-order Fermi acceleration. Here we have put the velocity dispersion $d\sigma^2_v$ in relation to the primordial density fluctuations $\sigma_p$ on a scale $R_p$, and note that their product is constant for the canonical case of a Harrison-Zeldovic spectrum, and we adopt the common convention $R_p=8h^{-1}\,$Mpc and $\sigma_p = \sigma_8\simeq0.8$.

Note that in Planck units, $\varsigma$ is simply the ratio of the radius of a structure $\tilde R$ to its mass $\tilde M$, or equivalently its inverse gravitational potential $\tilde \phi_G$ evaluated at $\tilde R$. Criticality is expressed by the exclamation mark, which relates it to the ratio of system temperature and particle mass, $T/m$, which is again equivalent to the velocity dispersion, which, as we have seen above, is roughly the same on all scales.\footnote{%
Counterrevolutionary readers who are not willing to believe this are invited to calculate $\varsigma$ for, e.g., the Sun, the Galaxy, the Virgo cluster, or other astrophysical systems that are in approximate thermal or virial equilibrium. 
}  
This relation is equivalent to setting the Jeans length of the structure, $\lambda_J$, equal to its size $R$, which defines the critical point at which structure formation proceeds in line with the expectations of self-organized criticality. The important role of the parameter $\varsigma$ is that it sets the magnitude for the conversion of gravitational to non-thermal energy on all scales.
 
\section{The energy reservoir of the universe}

\noindent Looking at a patch of mass $M$ and volume $V$ in the young, matter dominated Universe before structure formation became significant, we find in it four types of energy: The mass energy, $Mc^2$, the potential energy of the nearly homogeneous mass distribution in space, $U_M \simeq \frac35 Mc^2$, the radiative energy $\epsilon \ll M c^2$, and dark energy, $U_\Lambda = \Omega_\Lambda V \rho_c\,c^2$. The last two in the list are without significance here, thus we get a total energy budget of this patch of approximately $1.6\,M c^2$, which remains constant in comoving coordinates. All other energy in the Universe -- transversal or rotational motion, heat and thermal radiation, and of course also cosmic rays and their non-thermal offspring are bound to this budget, which is accessed through a series of processes ignited by structure formation. 

In the revolutionary sense, non-thermal energy is understood as a fraction of the potential energy released in gravitational collapse. From the Press-Schechter function (\ref{eq:PressSchechter}) we find for the power density released by gravitational collapse, averaged over a Hubble time,
\begin{equation}
\label{eq:Psi_nth}
\frac{d\,\bar\Psi_{\rm nth}}{d\ln M} = \frac{3 H_0 M_{\rm h} c^2}{5 \varsigma} \frac{dN_{\rm h}}{d\ln M_{\rm h}} \;\;=\;\; \frac{9\Omega_m\mu_M}{40 \pi\ln \widehat M_*}\,\frac{H_0^3 c^2}{G \varsigma} \gtrsim 10^{46} \;\mu_M\;\rm erg\,Mpc^{-3}\,yr^{-1}
\end{equation}
where we introduced the fraction $\mu_M \lesssim \Omega_b/\Omega_m \approx 0.1$ of matter which actually contributes to non-thermal energy conversion in the collapse. It is of no significance to the revolution whether this factor is applied to dark matter halo contraction with only the baryonic part of the matter contributing to non-thermal energy generation, or to smaller structures in which only baryonic matter collapses. It is instructive to compare this to the release of thermal energy in the nuclear burning of stars created during structure formation. This releases a fraction $\Omega_Z \langle b\rangle / \Omega_m m_p c^2$ of the total mass, with $\langle b\rangle \simeq 8\,$MeV being the average binding energy per nucleon and $\Omega_Z\approx 10^{-4}$ the density parameter for metals produced in structure formation. Distributing the released energy evenly over a Hubble time, we obtain an average thermal power density 
\begin{equation}
\label{eq:Psi_thermal}
\bar\Psi_{\rm th} \;\;=\;\; \frac{3}{8\pi}\,\frac{\langle b\rangle}{m_p}\,\frac{\Omega_Z H_0^3}{G} \;\;\approx\;\; 10^{49}\; \rm erg\,Mpc^{-3}\,yr^{-1}
\end{equation}
Integrating the non-thermal energy release (\ref{eq:Psi_nth}) over all mass scales removes the term $\ln\widehat M_* \approx 30$ from the denominator, thus $\bar \Psi_{\rm nth} \approx 10^{48}\rm erg\,Mpc^{-3}\,yr^{-1} \lesssim \bar\Psi_{\rm th}$. 

\section{Non-thermal energy conversion}

\noindent The next step of the revolution is to go from total energy to energy density. Let us first look at the collapse of a homogeneous cloud of (dark plus baryonic) matter. The conversion of gravitational potential energy density through our non-thermal cascade reads then  
\begin{equation}\label{nth-gc}
\label{eq:B_direct}
\frac{B_r^2}{8\pi} \;\;\approx\;\; \xi \left[r\frac{d}{dr}\frac{U(r)}{V(r)}\right]_{\tilde{r} \gtrsim \varsigma \tilde{M}} \;\lesssim\;\; \frac{9}{5 \pi}\,\frac{c^4}{G}\;\frac{\xi}{\varsigma^2}\,\frac{1}{r^2}
\end{equation}
where $U(r) = -5GM^2/3r$ and $V(r) = 4\pi r^3/3$ are the gravitational potential energy and the volume of the halo for the simplified assumption of a spherical homogeneous matter distribution at radius $r$, and we define $\xi = \xi_B\mu_M \approx 10^{-2}$ as the total efficiency for converting gravitational to non-thermal energy. It is striking to note that if we assume that the baryonic part of this structure further collapses into a black hole of mass $M_\bullet$ in a very fast process acting comparable to the free fall time of the system, we obtain 
\begin{equation} 
\label{eq:B_bh}
\frac{B_\bullet^2}{8\pi} \;\;\approx\;\; \frac{\xi_\bullet\, U(R)}{2\pi r_\bullet^2 [t_{\rm ff} v_{\rm ff}]_{r_\bullet}} \;\lesssim\;\; \frac{3}{5\pi^2}\,\frac{c^4}{G}\;\frac{\xi}{\sqrt{\varsigma^3\varsigma_\bullet}}\,\frac{1}{r_\bullet^2}
\end{equation}
with $[t_{\rm ff} v_{\rm ff}]_{r_\bullet}= \frac12 \pi R \sqrt{R/r_\bullet}$ is the product of free-fall time and free-fall velocity at a radius $r_\bullet = \varsigma_\bullet M_\bullet$ at which the conversion to non-thermal energy acts. This is typically the region at which part of the accreted energy is pulled into jets or winds which provide the right conditions for non-thermal processes, but as this is only a fraction of the baryonic part of the collapsing matter we have to assume $\mu_M \ll 0.1$ here (this also considers that only a fraction of all collapsing objects produce suitable jets or winds), thus we assume a total efficiency $\xi_\bullet \lesssim 10^{-3}$, but if we consider that $1\lesssim \varsigma_\bullet \lll \varsigma$ and put all the orders of magnitude together, we end up with $r_\bullet  B_\bullet \sim r\,B_r$. 

The relevant cases where such free-fall scenarios apply are stellar scales, $1 \lessapprox M/M_\odot \lesssim 300$, and they likely manifest themselves mostly in gamma-ray bursts or tidal disruption events. It may also have played a role during the formation of the seeds of super-massive black holes (SMBH) in the dark ages, although this is certainly not undisputed.\footnote{%
One of the really nasty properties of the dark ages is that they are dark. 
}
For larger SMBHs, we know that they are formed by subsequent mergers in structure formation. There is, however, a simple scaling between their masses and and those of the dark matter halos containing them, $
M_\bullet \approx 3\times 10^9 M_\odot (M_{\rm h}/M_*)^{3/2}$,
which is both predicted from simulations and confirmed by observations \cite{BS10}. This relation stems from the fact that halo mergers sooner or later lead to mergers of their central black holes, and the over-proportional growth is explained by accretion of matter, which is by order of magnitude limited by radiative pressure and yields   
\begin{equation}
\label{eq:B_edd}
\frac{B_{\rm edd}^2}{8\pi} \;\;\approx\;\; \frac{\xi\,L_{\rm edd}}{2\pi r_\bullet^2 v} \;\;\simeq\;\; \frac{3}{4\pi}\,\frac{c^4}{G}\;\frac{\xi \tilde M \tilde m_p \tilde m_e^2}{\alpha^2}\,\frac{1}{r_\bullet^2}
\end{equation}
where we have used $v\simeq c$ as it is reasonable for AGN jets, and $\alpha = e^2/\hbar c$ the fine structure constant. The dimensionless factor $3 \tilde M \tilde m_p \tilde m_e^2/ 2 \alpha^2 \simeq 3\times 10^{-22}\,M/M_\odot$ is essentially the Eddington luminosity in natural units. Note that for the largest SMBHs, $M_\bullet \lesssim 10^{10} M_\odot$, it is numerically comparable to $\varsigma^{-2}$ governing direct collapse processes, so for these mass scales we again obtain $r_\bullet  B_{\rm edd} \sim r\,B_r$.

\section{The Hillas-Plot mystery resolved}

\noindent From the bourgeois astroparticle physics point of view, the spooky legacy of the non-thermal spectre manifests itself in the famous\footnote{%
So famous that reproducing it is without relevance. Gentle readers are referred to their favorite www search engine.} 
``Hillas-plot'', which sets the magnetic field $B$ in a source of size $R$ in relation via 
\begin{equation}
\label{eq:E_Hillas}
\hat E_{\rm cr} \approx Z e \beta B R  \;\;\sim\;\; e B R \equiv E_H \sim 10^{20}\,{\rm eV}
\end{equation}
where $Z$ is s the charge number of the accelerated nucleus, $\beta = v/c$ a dimensionless characteristic velocity, and $E_H$ is commonly called the Hillas energy.\footnote{%
It is a bourgeois legend that this relation applies only to accelerators which need to confine their particles during acceleration. In fact it simply expresses that \textit{any} acceleration of charged particles requires effective electric fields ${\cal E}_{\rm eff} = |\mathbf{v\times B}|/c$, so that $E = Z\,e\,{\cal E}_{\rm eff} R$. Setting all geometric effects $\lessapprox 1$, this yields the Hillas expression. 
}
Originally thought just to give an empirical collection of astrophysical objects which could produce UHECR, the Hillas-Plot surprises us with the finding that (a) such objects seem to exist over 15 orders of magnitude in scale; (b) that they are all able to produce the required UHECR injection power $d\Psi_{\rm uhecr}/d\,\ln\,r \approx 10^{44}\;{\rm erg\,Mpc^{-3}\,year^{-1}}$, a number derived by Waxman {\cite{Waxman95}}; (c) that all sources with $\hat E_{\rm cr} \ll 10^{20}\,$eV have also $\Psi_{\rm uhecr} \ll 10^{44}\, \rm erg\,Mpc^{-3}\,yr^{-1}$; (d) that there exist no sources with $\hat E_{\rm cr} \gg 10^{20}\,$eV. This relation between maximum particle energy and injection power density, and its remarkable scale-invariance over 15 orders of magnitude, is the real mystery about the origin of cosmic rays.

It is main achievement of the revolution to bring down this mystery to the grounds of cosmological materialism. First, for all scenarios (\ref{eq:B_direct}), (\ref{eq:B_bh}), and (\ref{eq:B_edd}) we obtain
\begin{equation}
\label{eq:E_H_cosmo}
E_{\rm H} \equiv e B r \;\;\lesssim\;\; 3\,M_{\rm Pl} c^2 \frac{\sqrt{\alpha\xi}}{\phantom{\big[}\varsigma\phantom{\big]}} \sim  3\times 10^{19}\,\rm eV
\end{equation}
where $M_{\rm Pl} = \sqrt{\hbar c/G}$ is the Planck mass, and from (\ref{eq:Psi_nth}) we get
\begin{equation}
\label{eq:Psi_cr_cosmo}    
\frac{d\,\Psi_{\rm cr}}{d\,\ln r} \;\;=\;\; \xi\,\frac{d\,\bar\Psi_{\rm nth}}{d\ln M} \;\;\lessapprox\;\;  10^{44} {\rm \,erg\,Mpc^{-3}\,year^{-1}}\quad.
\end{equation}
Note that $d\Psi_{\rm cr}/d\,\ln r \propto \varsigma^{-1} \propto E_H$, and their ratio is only dependent on $\sqrt{\xi}$. We may regard the parameter $\xi$ the coupling constant of gravitational to electromagnetic forces on large scales, and note that $\xi\sim \alpha$, the electromagnetic coupling constant on quantum scales. The order of magnitude of both quantities, $d\Psi_{\rm cr}/d\,\ln r$ and $E_H$, is determined by $\varsigma$, the critical quantity of structure formation and thus driver of all non-thermal processes in the universe, which gets its value from the level of fluctuations present in the universe after inflation. This is the ``Hillas-plot'' mystery resolved: Both maximum observed cosmic ray energies and the universal cosmic ray flux result \textit{purely} from a combination of fundamental constants and cosmological parameters, on which they depend in almost the same way.   

\section*{The Hillas-Greisen-Zatsepin-Kuzmin (Gang-of-Four) conspiracy}

\noindent Another trail of phlegm left by the non-thermal haunting is that the two reasons given why the spectrum of cosmic rays must end, lead to the same order of magnitude: (a) The Hillas-energy of the sources, $\hat E_{\rm cr} \sim 10^{20}\,$eV, and (b) the Greisen-Zatsepin-Kuzmin cutoff, $\hat E_{\rm GZK} \lesssim m_p m_\pi c^2 / k T_{\rm cmb} \approx 10^{20}\,$eV. To shed light on the obvious conspiracy of this ``Gang-of-Four'' from the cosmological point of view, it is helpful to first rewrite the GZK energy as
\begin{equation}
\label{eq:E_GZK}
\hat E_{\rm GZK} \;\;\gtrapprox\;\; \frac{4 m_p c^2}{\alpha^2}\,\frac{m_\pi}{m_e} \frac{1+z_*}{1+z}\;,
\end{equation}
considering that the cosmic microwave background has been created at a redshift $z_*\simeq 1100$ with a temperature about an order of magnitude below the Rydberg energy. For the first time, the revolution now pays attention to cosmological epoch $(1+z)$, so before we proceed we have to note that $\varsigma \propto \Omega_m (1+z) + \Omega_\Lambda/(1+z)^2$. We then easily find
\begin{equation}
\label{eq:E_GZK_over_E_H}
\frac{E_{\rm GZK}}{E_{\rm H}} \;\;\gtrapprox\;\; \frac{2\,m_p m_\pi}{M_{\rm Pl}\,m_e}\;\frac{1+z_*}{\sqrt{\alpha^5\xi}}\; \left(\frac{c}{R_p \sigma_p H_0 \Omega_m^{0.3} }\right)^{\!\!2}
\left[1 + \frac{\Omega_\Lambda}{\Omega_m(1+z)^3}\right]\;\;\gtrapprox\;\; 1
\end{equation}
This does not really explain why the GZK cutoff is of the same order of magnitude as the Hillas energy, as this would require to understand how leptonic-mesonic-baryonic mass scales are related to primordial cosmological fluctuations.\footnote{%
Maybe inflation theory has an answer here, but this is currently beyond the revolutionary horizon.
}
It does show, however, that $E_{\rm GZK}\gtrapprox E_{\rm H}$ not only now, but for all redshifts. This is good and bad news for cosmogenic neutrinos, which rely on the common bourgeois assumption that $E_{\rm GZK} > E_{\rm H}$ at least at high redshifts. Bad news, because it means that expected fluxes further decrease, good news as it opens the possibility that the cosmogenic neutrino flux is dominated by a few exceptional sources sticking out of the average.\footnote{%
Unlike some other revolutions, the cosmological revolution can deal with exceptional individuals, and with it offers the bourgeoisie targets to satisfy their crave for smoking guns.
}

\section{Corollaries and fallacies about cosmic rays}

\noindent The revolution is generous. All arguments brought forward here apply only around order of magnitude. So what do we learn regarding the bourgeois questions about spectrum, composition and anisotropy of UHECR in the last decade? The answer is simple: Nothing! And this is good, because it is not the goal of the revolution to release us from understanding the details of cosmic ray production in the light of precise data.

What the revolution gives us instead is a universal view on cosmic rays, which does not lack predictive power either --- least relevant here is that it could predict cosmic ray properties in other universes. More important is that the overall extragalactic cosmic ray spectrum must be close to $dN/dE\propto E^{-2}$, mainly because the energy constraints are so tight that any steeper spectrum filling the entire extragalactic space would violate them. In turn this means that if evidence can be given for steeper spectra, that these could only pervade the Universe with a (very) limited filling factor and/or apply only to a (very) limited energy regime. Under no circumstances spectra steeper than $dN/dE \propto E^{-2.3}$ can be attributed to single power law sources of cosmic rays. 

The greatest gift of the revolution is, however, to expose a fatal fallacy in our thinking about cosmic rays, i.e., 
that the ubiquity of cosmic ray properties and the apparently extreme energies involved implies that only one exceptional source class\footnote{%
\ldots which may not even be regarded as known. It is one of the silliest prevarications of the bourgeoisie to explain mysteries with the unknown.
}
can be responsible for their production. Rather, the production of cosmic rays is revealed as an ubiquitous process spread over all scales of the Universe, and it is more likely than not that many source classes contribute which are all energetically connected through the criticality of structure formation. The existence of cosmic rays up to $10^{20}\,$eV is imprinted in this Universe and as normal as the existence of light.
    
As this fateful misconception is obviously rooted in the bourgeois desire for the supernatural (or at least, the beyond-current-physics), the \textit{CCCP} has decided in its infinite wisdom that after the revolution has triumphed, everybody who continues to use buzz-words like ``extraordinary'', ``enigmatic'',  ``puzzling'',  ``mysterious'' or similar in connection to UHECR will be excluded from funding (or relocated to research institutions in Siberia). 

\section{The utopia of cosmological astroparticle physics}

\noindent Unveiling a mystery is good, making predictions is better. And of course, the cosmological revolution has a utopia: the connection of astroparticle physics with the strongly evolving field of precision cosmology. In particular, the relation of UHECR production to the release of gravitational potential suggests a fruitful link to simulations of cosmological structure formation, which can give us information about the 4D distribution of cosmic ray sources on all scales, i.e., their formation, lifetime and evolution, as well as about the magnetic fields and large scale shocks governing UHECR propagation. In this way, the gross approach presented here may eventually lead to a theoretical framework for UHECR production which goes far beyond all current approaches in terms of predictive power and testability. 

On this base, the \textit{CCCP} will soon announce the five-year-plan for the final understanding of extragalactic cosmic rays and all related phenomena.

\begin{footnotesize}
\subsection*{Acknowledgements}
\noindent The author is indebted to the members of the ISSI International Team 323 for discussions during its meetings in 2014/15 when this idea was first conceived (http://www.issibern.ch/teams/bayesianmodel/), to the International Space Science Institute in Bern for hosting this team and offering generous financial support, and to Karl Marx and his followers around the world for the inspiration how to eventually publish it.

\end{footnotesize}

\end{document}